# A critical assessment of conformal prediction methods applied in binary classification settings

Damjan Krstajic[1§]

[1]Research Centre for Cheminformatics, Jasenova 7, 11030 Beograd, Serbia

[§]Corresponding author

Email address:

    DK: damjan.krstajic@rcc.org.rs

## Abstract

In recent years there has been an increase in the number of scientific papers that suggest using conformal predictions in drug discovery. We consider that some versions of conformal predictions applied in binary settings are embroiled in pitfalls, not obvious at first sight, and that it is important to inform the scientific community about them. In the paper we first introduce the general theory of conformal predictions and follow with an explanation of the versions currently dominant in drug discovery research today. Finally, we provide cases for their critical assessment in binary classification settings.

# Introduction

Conformal predictions (CP) is a very active research field in statistics and machine learning. It was introduced in Vovk *et al.* [1] and Sanders *et al.* [2] and further developed in Vovk *et al.* [3]. Since then several of its aspects have been developed [4]-[5], while in drug discovery it was pioneered by Norinder *et al.* [6]. As CP uses past experience to determine levels of confidence in new predictions, it is an approach with the potential of being useful in drug discovery projects where the issue of the application domain (AD) [7] is acute. Therefore, it is not surprising that in recent years there has been an increase in the number of scientific papers that suggest using CP in drug discovery [6],[8]-[15]. We consider that some of its versions applied in binary classification settings are embroiled in pitfalls, not obvious at first sight, and that it is important to inform the scientific community about them. In the paper we first introduce the general theory of CP and follow with an explanation of the versions of CP dominant in drug discovery research today. Finally, we provide the cases for their critical assessment in binary settings.

Damjan Krstajic [16] has published a criticism of a comparison between CP and QSAR applications. We will repeat his main point as it is applicable to binary classification models based on CP.

## Binary classification models

A binary classification statistical model is a predictive model F() that predicts a binary variable Y using values of variables X1,..,Xm. It can be viewed as the relationship Y=F(X1,,...,Xm). It is created using previously known values (Yi,Xi1,…Xim) i=1..N, which we refer to as the *training data*. As we are examining binary outputs, Y has only two values, which we shall refer to as positive and negative. The quality of the predictive model F() is measured by how well it predicts a previously unseen set of samples, which we refer to as the *test data*. There are various common measures of quality for binary classification models, such as misclassification error, specificity, sensitivity, etc. Furthermore, in binary classification settings it is common for a predictive model F() not only to predict whether something is positive or negative, but also to estimate its probability of being positive. The most common measure for assessing probabilistic classification models is the area under the ROC curve (AUROC).

## Conformal predictions in binary settings

Shafer and Vovk [17] designed CP for an on-line setting in which labels are predicted successively, each one being revealed before the next is predicted. This means that there is only one test sample. After the test sample is predicted, its true value is revealed and it is incorporated into the training set. The new predictive model is then built and ready to predict the next test sample. In such an environment they have defined several features of CP which makes it different from other statistical modelling approaches:

a) CP produces an N% *prediction region,* which contains possible predicted values with a probability of at least N%. In CP, binary classifiers may return the following four prediction regions for a single test sample: {positive}, {negative}, {positive and negative} and {null}. The last two predictions are usually referred to as 'both' and 'empty'.

b) CP requires a *nonconformity measure* to be specified. Given the nonconformity measure, the conformal algorithm produces a prediction region for any specified N%.

c) CP defines a new concept of *validity* for prediction with confidence. As CP is defined in an on-line setting, it is repeatedly applied to an accumulating data set, and not to independent data sets. Therefore, Shafer and Vovk [17] refer to an N% prediction region in the on-line method for binary classifiers as *valid* if N% of these predictions contain the correct label.

d) In addition to the validity of a method for producing N% prediction regions, Shafer and Vovk [17] discuss their efficiency. Shafer and Vovk [17] say that a prediction region is *efficient* if the prediction region is usually relatively small and therefore informative. In classification it is desirable to see an N% prediction region so small that it contains only a single predicted label, i.e. not 'both' nor 'empty'.

e) There are other features such as exchangeability and ability to be applied to all point estimates that make CP very attractive, but they are not relevant for our discussion.

We consider that the causes of problematic applications of CP in binary classification settings, which we will point to later in the text, arrise from a misunderstanding or misrepresentation of the above concepts.

**What does it mean in practice that a binary classifier provides a valid N% prediction region?**

In the binary classification setting, it does not mean that N% of the predictions will be correct, but that N% of the predictions will contain the correct label. This means that if we obtain 950 'both' predictions and 50 'empty' predictions from 1000 test samples, our predictions would be 95% valid, because each 'both' prediction contains the correct label. We do not see anything wrong with the definition of validity as such, but we question its practical value in the binary classification setting.

If our CP model produces an 86% valid prediction region it could be that we obtain, for example:

- 86% 'both' predictions and 14% 'empty'
- 50% 'both' , 36% correct, 10% false and 4% 'empty'
- 15% 'both, 71% correct, 2% false and 12% 'empty'
- 86% correct and 14% false single predictions

We see a dramatic difference in the practical value of the above cases all having the same valid 86% prediction region.

In a case when our conformal prediction model predicts all samples to be 'both' then we would have 100% validity, but in our view we would have 0% useful predictions. Therefore, we question the practical value of validity as a measure in binary classification settings.

**How can we choose the nonconformity measure?**

The nonconformity measure is the starting point for CP. Shafer and Vovk [17] define nonconformity measure as $d(\hat{z}(B), z)$, where z is a new example (test sample) and $\hat{z}(B)$ is a method for obtaining a point prediction $\hat{z}$ for a new example from a bag B of old examples (training samples).

Initially, Shafer and Vovk [17] provide distance to the neighbours for classification as an example of nonconformity measures in binary settings. Later, when they applied CP on Ronald A. Fisher's Iris dataset [18] they also used two other nonconformity measures: distance to the average of each species, and a support vector machine. Here we will only present the distance to the neighbours for classification as an example of a nonconformity measure $d(\hat{z}(B), z)$

$$d(\hat{z}(B), z) = \frac{distance\ of\ z's\ nearest\ neighbours \in B\ with\ the\ same\ label}{distance\ of\ z's\ nearest\ neighbours \in B\ with\ a\ different\ label}$$

As one can see in the above definition of their example of the nonconformity measure, Shafer and Vovk [17] use the values of the output binary variable Y in the training samples when calculating the nonconformity measure. We do not see anything wrong with the way Shafer and Vovk [17] have defined the nonconformity measure as such, but we would like to point out that it is not the same as the distance measures in AD [7], where only values of input variables are used, i.e. $(X_1,…X_m)$.

Shafer and Vovk [17] state that a nonconformity measure is a real-valued function which measures how different a test sample is from training samples. In some research fields outside of CP, such as AD [7], when someone "*measures how different a test sample is from training samples*" it is presumed that only values of input variables $\{X_1,…X_m\}$ are used for calculating the measure, while in CP that is not the case.

**Why efficiency is not measured in binary classification settings?**

As we mentioned earlier, when discussing efficiency in binary settings, Shafer and Vovk [17] point out that it is desirable to see an N% prediction region so small that it contains only a single predicted label, i.e. not 'both' nor 'empty'. We would like here to reiterate that Shafer and Vovk [17] designed CP for an on-line setting where there is only a single test sample. However, a variant of CP called Mondrian Conformal Predictions (MCP) [5] may be applied to a set of test samples. Furthermore, MCP is used in papers related to drug discovery, which we will analyse later.

When a CP approach in binary classification setting is applied to a set of test samples, we consider that it is important to know the percentage of single predictions (not 'both' nor 'empty') in a set of test samples. Our understanding of CP's terminology is that it would be the measure of efficiency, and we are puzzled as to why it is not introduced when a CP approach is applied to a set of test samples.

## Conformal predictions in drug discovery

Norinder *et al.* [6] introduced the use of CP in drug discovery research. We focus on their explanations and results, because they are the only authors, as far as we are aware, who explain the process of applying CP in binary settings properly and in full detail. Even though we are critical of their approach, we find their explanations to be clear and understandable. Furthermore, we would like to emphasize that our criticism does not suggest that there was any malicious intent on the part of the authors nor that they have deliberately misrepresented methods in any way. On the contrary, we think that there might be oversights on the part of the authors and that our criticism might in future help to improve their approach and its explanation.

There are, in our opinion, two major issues in Norinder *et al.* [6]. First, we question their choice of the nonconformity measure. Second, we demonstrate that their example is not useful in practice, and therefore not a good example for encouraging the use of CP.

Cortés-Ciriano and Bender [19] summarise the underlying concepts and practical applications of CP with a particular focus on drug discovery processes. Cortés-Ciriano and Bender [19] describe various versions of CP, and they list 28 drug discovery studies in which CP was implemented. Even though they describe current limitations in the field, our view is that Cortés-Ciriano and Bender [19] omitted to present major pitfalls and misapplications in the field. Furthermore, it seems to us that they misunderstood the validity in CP and, therefore, we believe misrepresented it unintentionally as a benefit of using CP. Here too our criticism does not entail any suggestion of malicious intent.

In our opinion, Norinder *et al.* [6] is a fundamental paper for discussing the use of CP in drug discovery, while Cortés-Ciriano and Bender [19] summarise the current state of CP applications in drug discovery.

**Choice of the nonconformity measure in Norinder *et al.* [6]**

For the binary classification case, Norinder *et al.* [6] defined the nonconformity score to be the probability for the prediction from the decision trees in the random forest, i.e. the score of a new compound is equal to the percentage of correct predictions given by the individual decision trees in the random forest. In other words, one would build a random forest model with Y as an output variable (e.g. solubility) and {X1,..,Xn} as input variables (e.g. chemical descriptors) and define the nonconformity score to be the probability for the prediction of Y from the decision trees in the forest. Therefore, authors presume that the measure of the difference between a test sample from training samples is equal to the percentage of correct predictions of Y (e.g. solubility) given by the individual decision trees in the forest.

They reference a paper by Devetyarov and Nouretdinov [20] for using such a nonconformity score. Devetyarov and Nouretdinov [20] mention three types of nonconformity measures of which the first one is equal to the percentage of correct predictions given for the sample by decision trees. However, the experiments and results in Devetyarov and Nouretdinov [20] are all for the other two types of nonconformity measures, which means apart from just being mentioned, Devetyarov and Nouretdinov [20] do not provide any practical justification for using the nonconformity measure as defined by Norinder *et al.* [6].

We are puzzled as to how predicting the probability of Y (e.g. solubility of a compound) may be a *a real-valued function which measures how different a test sample is from training samples*.

Cortés-Ciriano and Bender [19] report 18 published papers where the probability generated by a random forest is used as the nonconformity measure.

**Classification example in Norinder *et al.* [6]**

Norinder *et al.* [6] applied a variant of CP called Mondrian Conformal Predictions (MCP) [5]. In MCP, a training set is randomly divided into a proper training set and a calibration set. Norinder *et al.* [6] use a 70% (proper training set) and 30% (calibration set) split. The proper training set was used for model fitting, and the calibration set for constructing the prediction region.

In Figure 1 we show the predicted probabilities of classes A and B, which are exactly the same as Figure 1 in Norinder *et al.* [6]. They are an example of results on a calibration set consisting of 21 compounds that authors used for explaining how the prediction region is created.

| class A | class B |
|---------|---------|
| 0.002   | 0.01    |
| 0.15    | 0.08    |
| 0.23    | 0.21    |
| 0.40    | 0.36    |
| 0.48    | 0.43    |
| 0.70    | 0.51    |
| 0.75    | 0.64    |
| 0.80    | 0.72    |
| 0.95    | 0.75    |
| 0.98    | 0.80    |
|         | 0.95    |

Figure 1.

We are not questioning their explanation of the way the prediction region is created, but rather their choice of an example and its consequences. As we are dealing with a binary classification, let's say, for ease of presentation, that 'class A' is a negative class and 'class B' positive. We calculated AUROC for 21 predicted probabilities on the calibration set and found it out to be 0.527. Furthermore, if we take 0.5 to be a threshold for predicting labels, then accuracy of predictions is 0.524. We doubt that anybody would use such a model in practice.

We believe that presenting an example with almost random predictions is an oversight on the part of the authors, and we would not have mentioned had it not been the only proper example amongst published papers where CP is applied in drug discovery. If an example with good predictions was presented, instead of an example with random predictions, then it would be obvious that a number of otherwise correct predictions (in the binary sense) would become 'both' or 'empty', and therefore show a weakness of the CP approach.

We would like to point out that Norinder *et al.* [6] do not inform the readers regarding AUROC nor of the accuracy of their predictions on the calibration set. Unfortunately, the practice of not reporting the performance on the calibration set has been accepted by other authors. We are not aware of any published paper where CP is applied in drug discovery and where the performance on the calibration set is presented.

# Issues when comparing CP with binary predictions

In binary classification models there are 2 possible prediction outcomes {*positive, negative*}, while in CP there are 4 possible prediction regions {*positive, negative, 'both', 'empty'*}. As we have described earlier, one may calculate, for example, the misclassification error, specificity and sensitivity of a binary classification model. However, which error statistics may one use for prediction regions generated by a CP model?

### 'both' as correctly classified

Bosc *et al.* [15] described a test study that directly compares CP with binary classification models in a QSAR setting. Their approach treats 'empty' prediction regions as false predictions, while for 'both' prediction regions they analyse cases when 'both' is considered correctly classified as well as when 'both' is treated as a false prediction.

Damjan Krstajic [16] has published a criticism of the approach presented by Bosc *et al.* [15]. Here we will only repeat the main point of his criticism which Bosc *et al.* [21] omitted to comment upon in their reply. Half of the comparisons between CP and QSAR presented in Bosc *et al.* [15] examine situations when predictions assigned to 'both' are considered correctly classified. Thus, if a sample has a positive output value and it is predicted as 'both' it would be treated as correctly predicted. Furthermore, if it has a negative output value it would again be treated as correctly predicted. This implies that if we have a CP model with all 'both' predictions we would have 100% correct predictions. In our opinion, this does not make sense. How can someone in practice transform 'both' predictions into correct classifications?

**Limiting the number of false positives**

In our opinion, Cortés-Ciriano and Bender [19] misunderstood the validity in CP and unintentionally misrepresented it as the benefit of using CP. In their introduction section they state: "*The prediction of well-calibrated (or valid) confidence regions by CP, which is usually not the case for other modelling methods, guarantees a lower bound for validation rates, and permits to limit the number of false positives, thus increasing the retrieval rate of active compounds in preclinical drug discovery.*". They reference Norinder *et al.* [6] for such a statement, but we could not find that Norinder *et al.* [6] suggested that CP limits the number of false positives.

False positives are negative samples predicted as positive. We don't see how a 95% valid prediction region in CP can limit the number of false positives. How did we limit the number of false positives if we obtained 950 'both' predictions and 50 'empty' predictions from 1000 test samples?

**Quality of single predictions**

In our opinion, a potential benefit of using CP is in the case where test samples with single CP prediction regions have better statistics than all predictions of test samples using standard binary classification models.

# Discussion

We would like to reiterate that we are not criticising the CP theory, but its presentations of benefits. We are not questioning the results of any authors, but their scientific value. There are some important details in the CP theory that have implications different from what some authors present them to be. Here we will summarise them.

1) An N% prediction region for a binary classifier is *valid* if N% of these predictions contain the correct label. Saying that N% of predictions contain the correct label without explaining that the 'both' prediction, i.e. {positive, negative), contains the correct label but that it is not actually the correct label, might lead readers to misunderstand the true meaning of the validity in CP. We would like to point out that Shafer and Vovk [17] clearly and fully explain the pros and cons of the validity measure in CP.

2) Shafer and Vovk [17] say that a nonconformity measure is a real-valued function that measures how different a test sample is from training samples. In our opinion, there is a need for more clarification here. As we have shown, in the examples that Shafer and Vovk [17] present, the calculation of nonconformity measures presumes the use of the output binary variable Y, as well as input variables $X_1,..,X_m$.

3) Cortés-Ciriano and Bender [19] present 18 published papers which use the percentage of correct predictions given by the individual decision trees in a random forest as the measure of how different a test sample is from training samples? We cannot find any theoretical or empirical evidence which would support using the percentage of correct predictions as the measure of how different a test sample is from training samples.

4) In our opinion, there is still an unresolved problem in CP as to how to deal with 'both' and 'empty' prediction regions. Bosc *et al.* [15] presume that 'both' predictions may be treated as correct classification. We think that such practice is not logical. How can someone in practice treat a {positive, negative} prediction region, i.e. 'both' prediction, as a correct classification? How can it be useful in science to examine situations in which we assume that we know something which we cannot know? We are puzzled as to how this practice is accepted in the scientific community.

# Conclusion

We have presented here our critical assessment of CP methods applied in binary classification settings. We would like to point out that we do not have anything against any of the authors whose methods we have criticised here. Our intention is to inform the scientific community of a different view, which is currently not present.

# Acknowledgments

The author would like to thank his mother, Linda Louise Woodall Krstajic, for correcting English typos and language improvements in the text.

# Declarations

**Competing interests**
The author declares that he has no competing interests.

**Funding**
No funding received.